\documentclass[prl, showpacs, superscriptaddress, twocolumn, floatfix]{revtex4}
\usepackage{graphicx,color,amssymb}

\date{11/27/10}

\begin{document} 

\title{Enhancing the Coherence of a Spin Qubit by Operating it as a  
Feedback Loop Controlling its Nuclear Spin Bath}

\author{Hendrik Bluhm}  
\thanks{These authors contributed equally to this work.}
\affiliation{Department of Physics, Harvard University, Cambridge, MA 02138, USA}
\author{Sandra Foletti}
\thanks{These authors contributed equally to this work.}
\affiliation{Department of Physics, Harvard University, Cambridge, MA 02138, USA}
\author{Diana Mahalu}
\affiliation{Braun Center for Submicron Research, 
Department of Condensed Matter Physics, 
Weizmann Institute of Science, Rehovot 76100, Israel}
\author{Vladimir Umansky}
\affiliation{Braun Center for Submicron Research, 
Department of Condensed Matter Physics, Weizmann Institute of Science, 
Rehovot 76100, Israel}
\author{Amir Yacoby}
\email{yacoby@physics.harvard.edu}
\affiliation{Department of Physics, Harvard University, Cambridge, MA 02138, USA}

\begin{abstract}
In many realizations of electron spin qubits the dominant source of
decoherence is the fluctuating nuclear spin bath of the host material.
The slowness of this bath lends itself to a promising mitigation
strategy where the nuclear spin bath is prepared in a narrowed state
with suppressed fluctuations. Here, this approach is realized for a
two-electron spin qubit in a GaAs double quantum dot and a nearly
ten-fold increase in the inhomogeneous dephasing time $T_2^*$ is
demonstrated. Between subsequent measurements, the bath is prepared by
using the qubit as a feedback loop that first measures 
its nuclear environment by coherent precession, and then
polarizes it depending on the final state. This
procedure results in a stable fixed point at a nonzero 
polarization gradient between the two dots, which 
enables fast universal qubit control.
\end{abstract}


\maketitle

Spins in semiconductors are attractive qubits because of their long
coherence times \cite{Ladd2005, Tyryshkin2003, Childress2006}, their
electrical control and readout \cite{Petta2005}, and their potential
for scalability \cite{Taylor2005}.  Few-electron quantum dot devices
have been used successfully in recent years to demonstrate universal
control of electron spin qubits as well as single shot readout
\cite{Elzerman2005, Press2008, Barthel2009, Foletti2009}.  However,
interaction of the qubit spin(s) with nearby nuclear spins is a
significant source of decoherence in several systems \cite{Petta2005,
Childress2006, Koppens2006, Fischer2008, Tyryshkin2003}.  It is
therefore very attractive to prepare the spin environment of the
electron in a way that mitigates this decoherence.  One approach would
be to polarize it \cite{Takahashi2008}, but the nearly complete
polarization required for improved coherence\cite{Coish2004}
is difficult to achieve. 

Here, we present a method to narrow the
distribution of the fluctuating nuclear hyperfine field
while maintaining a weak polarization. In addition to the reduced
decoherence, such a narrowed state is of interest for studying the
long-time quantum dynamics arising from the spin-bath interaction
\cite{Coish2008}.  
Our method relies on first letting the qubit evolve under the
influence of the bath. The resulting final state of the qubit controls
the effectiveness of a subsequent dynamic nuclear polarization step.
The qubit thus acts as a complete feedback loop, and the outcome of
its measurement of the controlled variable does not need to be known
to the outside world.

Electron-nuclear feedback mechanisms were previously observed in
resonance locking experiments under microwave \cite{Vink2009} and
optical irradiation \cite{Xu2009, Latta2009, Greilich2007}.  In
Refs. \cite{Vink2009, Latta2009}, a narrowing of the hyperfine field
was inferred from the observed bidirectional polarization keeping the
system on resonance, but not experimentally verified. In
Ref. \cite{Xu2009}, narrowing was detected spectroscopically, 
and Ref. \cite{Greilich2007}  studied an ensemble of 
optically controlled quantum dots. Here, we
directly measure the narrowed distribution of the hyperfine field and
the dephasing time, $T_2^*$, of a single, electrically controlled qubit. 
$T_2^*$ is enhanced by nearly an
order of magnitude. In contrast to previous experiments, where the
feedback mechanism is intrinsic to the polarization dynamics
\cite{Danon2009, Rudner2007}, we have intentionally designed it by
manipulating the qubit. 

The spin qubit studied in this work employs the $m=0$ subspace of two
electron spins in a double quantum dot. The energy splitting between
the two basis states depends on the hyperfine field gradient between
the two dots. Ref. \cite{Reilly2008C} reported a complete elimination
of this gradient and an associated enhancement of $T_2^*$ in similar
devices.  However, there is now a more likely
interpretation of that experiment in terms of a loss of readout
contrast due to a {\em large} hyperfine field gradient 
that accelerates inelastic decay \cite{Barthel2010}.
Furthermore, maintaining a nonzero average field gradient as done here
is essential for universal fast electrical control of the qubit
\cite{Foletti2009}.

\begin{figure}
\includegraphics[width=8.6cm]{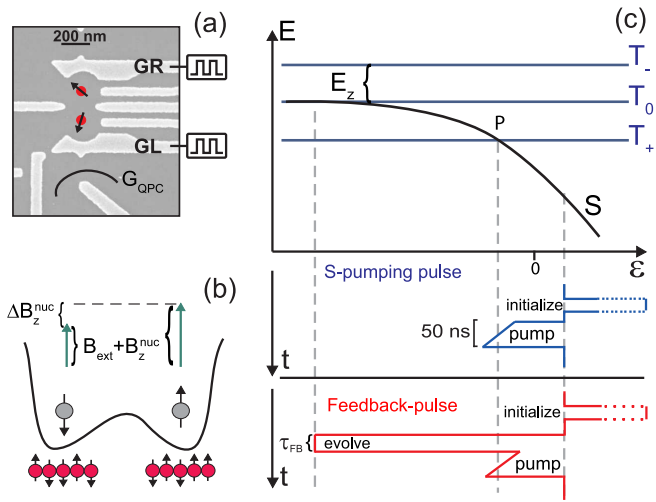}
\caption{\label{fig:schem}(color online)
Basic principles of the experiment. (a) 
Scanning electron micrograph of a similar device.
The qubit is read out by measuring the conductance $
G_{QPC}$ of the quantum point contact.
(b) Illustration of the dynamics for fully separated electrons. 
The energy splitting between $|\uparrow\downarrow\rangle$ and
$|\downarrow\uparrow\rangle$ is proportional to the hyperfine field 
gradient across the two dots. This level splitting results in 
coherent oscillations between $|S\rangle$ and $|T_0\rangle$.
(c) Top: 
Energies of the relevant spin states as a function of the detuning
$\varepsilon$, which can be controlled with nanosecond time resolution
by applying voltage pulses to gates GL and GR.  Bottom: Schematic of
$\varepsilon(t)$ for the $S$ and feedback pump pulses.  The total
duration of a single pulse is 250 ns.}
\end{figure}

The double quantum dot forming our qubit is created by locally
depleting a 90 nm deep two dimensional electron gas (2DEG) with
electrostatic gates [Fig. \ref{fig:schem}(a)].  Each dot is tunnel
coupled to a lead and the inter-dot tunnel coupling is $t_c \sim 20
\mu$eV.  The phase space of our qubit is spanned by
the singlet $|S\rangle \equiv (|\uparrow \downarrow\rangle -
|\downarrow \uparrow\rangle)/\sqrt 2$ and the triplet 
$|T_0\rangle \equiv (|\uparrow \downarrow\rangle + |\downarrow
\uparrow\rangle)/\sqrt 2$. The remaining two states, $|T_+\rangle \equiv
|\uparrow \uparrow\rangle$ and $|T_-\rangle \equiv |\downarrow
\downarrow\rangle$, are split off by the Zeeman energy $E_Z = g^*
\mu_B B_{ext}$ induced by an external magnetic field $B_{ext}$.
Throughout this work, $B_{ext}$
= 0.7 T was applied along the $z$-axis, parallel to the 2DEG.
The encoding of the qubit in two spins enables fast electrical control
via the energy difference, $\varepsilon$, between states with both
electrons in one dot and one electron in each dot, respectively
\cite{Levy2002, Petta2005, Foletti2009}.  The energies of the four
spin states depend on $\varepsilon$ as shown in
Fig. \ref{fig:schem}(c).  High frequency coaxial lines connected to
two gates GL and GR allow rapid changes of $\varepsilon$.

\begin{figure}
\includegraphics{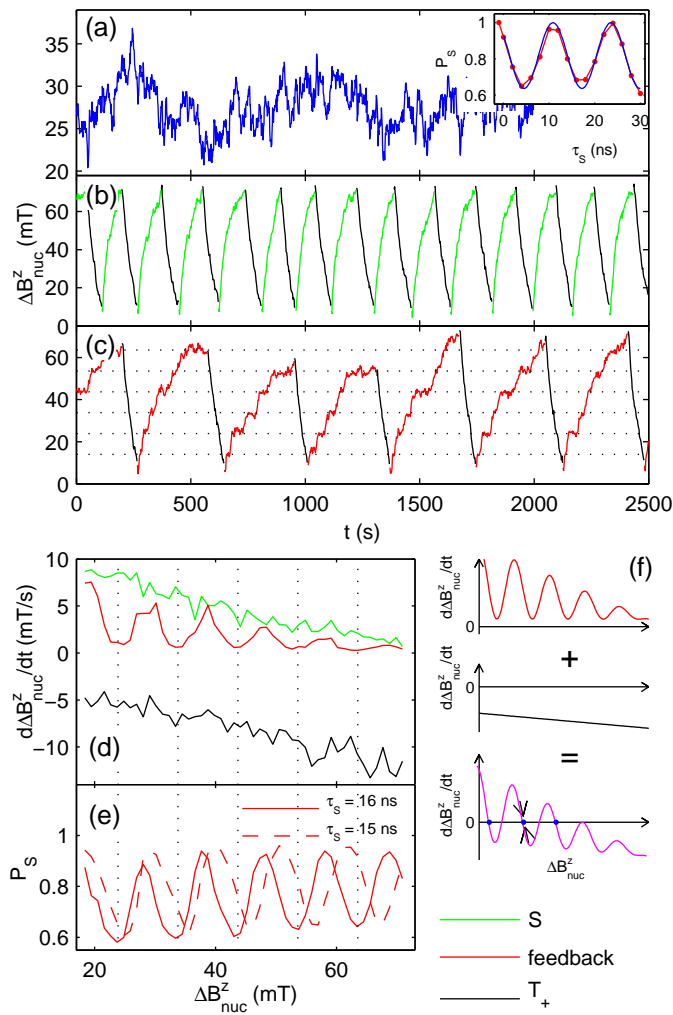}
\caption{\label{fig:pump}(color online) Polarization rates.  (a) Time
trace of $\Delta B_{nuc}^z$ showing the fluctuations of the nuclear
hyperfine field. The nonzero mean of $\Delta B_{nuc}^z$ is due to
pumping (without feedback) between measurements. Each data point
reflects a fit to the $S-T_0$ precession rate as shown in the inset.
(b) Same, but alternating between 0.4 s $S$ and $T_+$-pumping between
measurements. The legend at the lower right is valid for panels 
(b)-(f).  (c) Same for a feedback pulse with $\tau_{FB}$ = 15 ns and
$T_+$.  (d) Pump rates extracted by averaging over time traces such as
those in panels (b) and (c), but much longer.  $d\Delta B_{nuc}^z/dt$
is normalized by the time spent pumping.  (e) Singlet probability
$P_S$ after a fixed mixing time $\tau_S$ as a function of $\Delta
B_{nuc}^z$, extracted from the same data as the red curves in panels
(c) and (d) \cite{aux}.  $P_S$ is normalized by the change of the
charge signal when sweeping gate voltages across the charge transition
\cite{Foletti2009}. It does not decay all the way to zero because of
inelastic decay of the metastable $T_0$ during the measurement
process.  The dotted lines in panels (c)-(e) indicate where the
polarization stagnates because the qubit is swept through the
$S$-$T_+$ transition in $|T_0\rangle$. 
The best match between $P_S$ and $d\Delta B_{nuc}^z/dt$ is obtained
for $\tau_S$ = 16 ns, whereas $\tau_{FB}$ = 15 ns.
This difference probably reflects finite pulse rise times. 
 (f) Schematic illustration of the summing of the pump rates as
shown in panel (d) from $S$ (top) and $T_+$-pumping (middle) to obtain
stable fixed points (blue dots, bottom).  Fluctuations of $\Delta
B_{nuc}^z$ away from these points are compensated by a restoring pump
effect (arrows).}
\end{figure}

At $\varepsilon \ll 0$, the electrons are fully separated 
and acquire a phase at a
rate proportional to the magnetic field in each dot
[Fig. \ref{fig:schem}(b)].  These local fields are the sum of the
homogeneous external field and a hyperfine field proportional to the
nuclear polarization parallel to $\mathbf{B}_{ext}$.  Any difference
$\Delta B_{nuc}^z$ between the two hyperfine fields leads to coherent
oscillations of the qubit's state between $|S\rangle$ and
$|T_0\rangle$.
We probe the hyperfine field gradient, $\Delta B_{nuc}^z$, by
measuring this free precession using a standard
prepare-evolve-measure cycle \cite{Petta2005, Barthel2009, Foletti2009}
relying on spin to charge conversion and a quantum point contact for readout.  
The probability $P_S$ of finding
the electron in $|S\rangle$ varies sinusoidally with the
evolution time $\tau_S$ [inset to Fig. \ref{fig:pump}(a)], with a
frequency given by $f = |g^* \mu_B\Delta B_{nuc}^z|/h$, 
where $g^*\approx -0.4$ is the $g$-factor for a confined electron in 
GaAs. Fitting a sine
curve every time a $\tau_S$ sweep is completed yields a time trace of
$|\Delta B_{nuc}^z|$ with a sampling rate of about 1 Hz \cite{aux},
which is fast enough to probe slow random variations of the nuclear
polarization \cite{Reilly2008A}.

This near real-time measurement of $\Delta B_{nuc}^z$ allows us to use
pump cycles discussed in detail in Refs. 
\cite{Foletti2009, Petta2008, Reilly2008C} to compensate the fluctuations 
of the spin bath.  These pump cycles use the
degeneracy point of $|S\rangle$ and $|T_+\rangle$ 
[point P in Fig. \ref{fig:schem}(c)] to
exchange spin between the electrons and the nuclei.
In the so called $S$ ($T_+$) pumping cycle, we prepare a $|S\rangle$
($|T_+\rangle$) and then sweep trough the $S$-$T_+$ transition, which
builds up a polarization of the same (opposite) sign
as the applied field in each of the dots.  
How these pump cycles affect $\Delta B_{nuc}^z$
depends on the imbalance of the polarization rates in the two dots, 
which may, for example, arise from different dot sizes
\cite{Gullans2010} due to disorder.  Experimentally, we can probe
their effect by running on the order of $10^6$ cycles (at a 4 MHz
repetition rate) between measurements of $\Delta B_{nuc}^z$. 
This alternation between measuring and pumping was applied throughout
the remainder of the paper. We find
that the two pump cycles always change $\Delta B_{nuc}^z$ in opposite
directions.  If the gradient reaches zero while pumping, it
immediately increases again, which suggests a sign change. This
behavior is consistent with 
Ref. \cite{Foletti2009}, where the $T_+$-cycle was first introduced.

\begin{figure}
\includegraphics{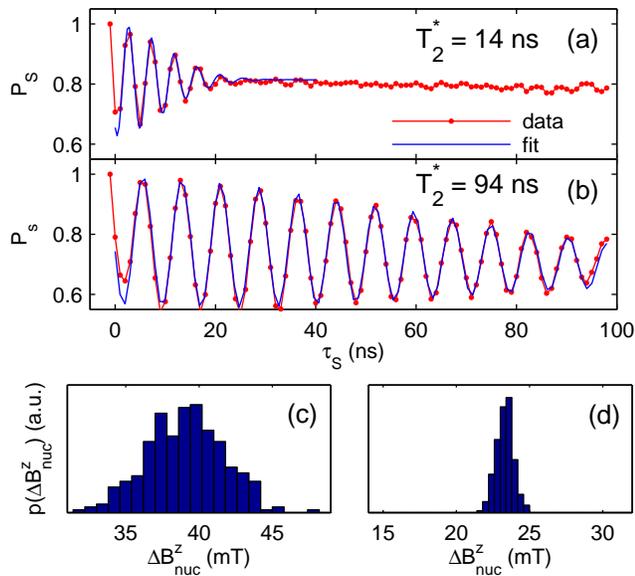}
\caption{\label{fig:T2star}(color online)
$T_2^*$ enhancement.  (a) Ensemble
average of $S$-$T_0$ oscillations for fixed pumping (0.7 s
$T^+$-pumping and 0.25 s $S$-pumping repeated every 2.2 s).  The
pumping maintains a nonzero $\Delta B_{nuc}^z$ = 39 mT whose 
fluctuations lead
to a dephasing time of $T_2^*$ = 14 ns. Other pump times give
different oscillation frequencies, but similar dephasing times.  
$P_S$ is normalized as explained in the caption of Fig. \ref{fig:pump}.
(b) Same measurement, but for 30 ms of feedback pumping with
$\tau_{FB}$ = 30 ns and 31 ms of $T_+$-pumping after each 39 ms long
measurement. The feedback extends $T_2^*$ to 94 ns.  The blue lines
show the fits to $1 - P_S \propto 1-\cos(g^* \mu_B \Delta B_{nuc}^z
\tau_S/h + \delta) e^{-(\tau_S/T_2^*)^2} + \alpha \tau_S$ used to extract
$T_2^*$ and $\Delta B_{nuc}^z$ = 23 mT.  
The linear term in the fit model accounts for crosstalk of
the gate pulses to the charge sensor,
and $\delta$ corrects for finite pulse rise times.
(c), (d) Corresponding distributions of $\Delta B_{nuc}^z$, 
obtained by histogramming instantaneous values of 
 $\Delta B_{nuc}^z$ before ensemble averaging without (c) and with 
pulse feedback (d).
}
\end{figure}

To quantify the effect of pumping, we switch between $S$
and $T_+$-pumping whenever $\Delta B_{nuc}^z$ reaches one of
two predetermined limits. This leads to a saw-tooth like
time dependence of $\Delta B_{nuc}^z$,  as shown in Fig. \ref{fig:pump}(b). 
Averaging over many such cycles \cite{aux}
yields the mean rate of change of the gradient, $\Delta B_{nuc}^z/dt$,
as a function of its value $\Delta B_{nuc}^z$
[Fig. \ref{fig:pump}(d)].  The approximately linear relation 
for $S$ and $T_+$-pumping  reflects the relaxation of the polarization 
due to spin diffusion. For a fixed pump time and pulse,
$\Delta B_{nuc}^z$ saturates once pumping and relaxation balance each other,
but continues to fluctuate on time scales of up to minutes 
[Fig. \ref{fig:pump}(a)], with an
rms-amplitude $\delta \Delta B_{nuc}^z$ of about 3 mT
[Fig. \ref{fig:T2star}(c)]. These fluctuations lead to a Gaussian decay
of coherent $S$-$T_0$ oscillations after a time $T_2^* = \hbar\sqrt 2/
(g^* \mu_B \delta \Delta B_{nuc}^z) = 14$ ns when averaging
over many $\tau_S$ sweeps with different oscillation frequencies
[Fig. \ref{fig:T2star}(a)]. 

Pumping with the $S$ and $T_+$-pulse for a fixed time generally does
not change $\delta \Delta B_{nuc}^z$ appreciably. However, our ability
to rapidly measure and manipulate $\Delta B_{nuc}^z$ enables us 
to narrow its distribution using software based feedback.
A proportional-integral feedback loop 
determines the type and duration of pumping between measurements
from the time trace of $\Delta B_{nuc}^z$.
This procedure reduced 
$\delta \Delta B_{nuc}^z$ by about a factor of 2, corresponding to
$T_2^* \approx$ 30 ns. It was limited by the $\approx 1$ Hz
sampling rate of $\Delta B_{nuc}^z$ and could thus be improved with
a faster readout technique \cite{Reilly2007, Barthel2009}.

While this software feedback method already uses the same qubit to
measure and polarize the nuclei, these two tasks are linked via a
relatively slow readout process and the measurement computer.  In
order to speed up the feedback response, we have
bypassed this connection by combining both operations into a single
pulse derived from the $S$-cycle [Fig. \ref{fig:schem}(c) bottom]. 
In such a pulse feedback cycle, the
qubit first probes its nuclear environment
and then polarizes it depending on the result: after initialization in
$|S\rangle$, the qubit is allowed to evolve at $\varepsilon \ll 0$ for
a time $\tau_{FB}$.  As in the probe
cycle, its state oscillates between $|S\rangle$ and $|T_0\rangle$,
and the probability of ending in $|S\rangle$ is given by $(1+\cos(g^*
\mu_B \Delta B_{nuc}^z \tau_{FB}/\hbar))/2$.  Upon sweeping
$\varepsilon$ past the $S$-$T_+$ transition, a nuclear spin can only
be flipped by the $|S\rangle$-component of the qubit's state emerging
from the evolution.  Thus, the pump rate should be
proportional to the $\Delta B_{nuc}^z$-dependent singlet probability
when averaged over many cycles.  We have verified this behavior by
characterizing the feedback-pulse in the same way as the $S$ and
$T_+$-pulses.  The measured mean pump rate,
$d\Delta B_{nuc}^z/dt$, oscillates as a function of $\Delta B_{nuc}^z$
between 0 and the value corresponding to $S$-pumping
[Fig. \ref{fig:pump}(d)]. As expected, this modulation follows the
singlet probability, $P_S$, extracted from the same data
[Fig. \ref{fig:pump}(e)]. Its period is given by $h/g^*\mu_B
\tau_{FB}$.

In order to obtain a stable fixed point for $\Delta B_{nuc}^z$, the
pump rate has to cross zero with a negative slope. Fluctuations
of $\Delta B_{nuc}^z$ away from the fixed point are then corrected by
an opposing pump effect [Fig. \ref{fig:pump}(f) bottom].  
However, the feedback cycle alone pumps
nuclei in one direction only [Fig. \ref{fig:pump}(d),(f) top]. 
At the minima of $P_S$, $d\Delta B_{nuc}^z/dt$ approaches zero because
the qubit is swept past the $S$-$T_+$ transition in a
$|T_0\rangle$. This stagnation of the polarization results in
steplike structures in the time traces in Fig. \ref{fig:pump}(c). On
either side of these points, $d\Delta B_{nuc}^z/dt$ remains positive.
Thus, $\Delta B_{nuc}^z$ continues to grow once a fluctuation pushes
it past one of these unstable fixed points \cite{aux}.  The required sign
change can be engineered by preceding the feedback pulse with
some amount of $T_+$-pumping \cite{aux}. The resulting mean pump rate is the sum of a weakly 
$\Delta B_{nuc}^z$-dependent
negative  $T_+$-pulse component and a positive contribution
oscillating with $\Delta B_{nuc}^z$ from the feedback pulse [
Fig. \ref{fig:pump}(f)].

To test the stabilizing effect of this pulse combination, 
we  applied it for a fixed time between measurements of
$P_S$ for different $\tau_S$. 
Figs. \ref{fig:T2star}(b),(d) demonstrate an 
enhancement of $T_2^*$ from 14 ns to 94 ns, and the corresponding
narrowing of the distribution of $\Delta B_{nuc}^z$ around
a fixed point.
Here, the pump pulses were applied for 61 ms per 100 ms interval.
The remaining 39 ms were spent measuring $P_S$ for a single $\tau_S$, 
and the data were averaged over 232 sweeps of $\tau_S$.
We estimate that the pump rate achieved here limits
the feedback response time to  $\approx$1 s.

In a quantum processor, the  measurements could be replaced by a
sequence of gate operations, whose fidelity would be substantially
improved by the reduction of fluctuations. 
For a 1 ns $\pi$-rotation generated by a gradient of 100
mT \cite{Foletti2009} with rms fluctuations of $\delta \Delta
B_{nuc}^z$ = 0.5 mT as demonstrated here, the fidelity
is $\pi^2 {\delta \Delta B_{nuc}^z}^2/4 {\Delta B_{nuc}^z}^2 \lesssim
10^{-4}$ \cite{Vandersypen2004}.  However, due to the slowness of the
nuclear bath, an error of order unity accumulates after only 100 such
gates.  This limitation could be overcome by 
making gates insensitive to $\delta \Delta B_{nuc}^z$ to first order
\cite{Khodjasteh:DECgates}.  A pulse angle error of order ${\delta
\Delta B_{nuc}^z}^2/{\Delta B_{nuc}^z}^2 \sim 10^{-4}$ would allow
$10^4$ operations per error. In either case, the improvement in gate
fidelity is at least quadratic in the narrowing ratio. Thus, narrowing
procedures are very effective at overcoming the limitations imposed by
a fluctuating nuclear spin bath. The flexibility of our approach should
allow an adaptation to other systems or other protocols on the same
system.

\acknowledgments{This work was supported by ARO/IARPA, the Department
of Defense and the National Science Foundation under Award No.
0653336. This work was performed in part at the Center for Nanoscale
Systems (CNS), a member of the National Nanotechnology Infrastructure
Network (NNIN), which is supported by the National Science Foundation
under NSF Award No. ECS-0335765.}

\end{document}